\documentclass[twocolumn,showpacs,prb,amsmath,amssymb,floatfix]{revtex4}
\usepackage{graphicx}     
\begin{document}

\title{Dynamical mean field theory of the Gutzwiller-projected
BCS Hamiltonian: Phase fluctuations and the pseudogap}
\author{K. Park}
\affiliation{School of Physics, Korea Institute for Advanced
Study, Seoul 130-722, Korea}
\date{\today}

\begin{abstract}
One of the most prominent problems in high temperature
superconductivity is the nature of the pseudogap phase in
underdoped regimes; particularly important is the role of phase
fluctuations. The Gutzwiller-projected BCS Hamiltonian is a useful
model for high temperature superconductivity due to an exact
mapping to the Heisenberg model at half filling and generally a
very close connection to the $t$-$J$ model at moderate doping. We
develop the dynamical mean field theory for the $d$-wave BCS
Hamiltonian with on-site repulsive interaction, $U$, physically
imposing the partial Gutzwiller projection. For results, two
pseudogap energy scales are identified: one associated with the
bare pairing gap for the singlet formation and the other with the
local phase coherence. The real superconducting gap determined
from sharp coherence peaks in the density of states shows strong
renormalization from the bare value due to $U$.
\end{abstract}

\maketitle

\section{Introduction}
\label{introduction}

Among the most fundamental issues in high temperature
superconductivity is the pairing mechanism. However, even without
its specific knowledge, many properties of the superconducting
phase can be well understood in the framework of the conventional
BCS theory with modification of the $d$-wave pairing symmetry.
On the other hand, there are a variety of peculiar behaviors
observed in cuprates, which cannot be explained within the
conventional BCS theory. These unconventional behaviors are
pronounced in underdoped regimes where superconducting order is
subject to strong phase fluctuations. While experimental setups
are diverse, these unconventional behaviors can be attributed to
the existence of an energy gap called the pseudogap
\cite{LeeNagaosaWen,Timusk}. Complete understanding of high
temperature superconductivity, therefore, requires a consistent
theoretical framework which not only provides an explanation for
pairing, but also contains the pseudogap phenomenon as a natural
part.

Experiments in fact suggest two distinctive pseudogaps
\cite{LeeNagaosaWen}. In the first class of experiments, the
pseudogap phenomenon can be understood in terms of the formation
of spin singlet pairs. Since spin singlet states are rigid against
spin flip, it is expected that, if there remains a tendency toward
the singlet formation even above $T_c$, the spin susceptibility
should be reduced from that of the paramagnetic phase, which is
indeed consistent with Knight-shift measurements \cite{Curro}.
Also, the decrease in the specific heat \cite{Loram} can be
understood in terms of the spin entropy loss. Furthermore, the
high temperature pullback of the leading edge observed in
angle-resolved photoemission spectroscopy \cite{ARPES} can be
interpreted as a consequence of the energy cost in breaking pairs.
The gap formation in the frequency-dependent $c$-axis conductivity
\cite{caxis} can be also explained similarly.

In the second class of experiments, the pseudogap phenomenon is
understood so that, despite the absence of global coherence, the
local phase is well defined. The main experimental tool in this
class is the Nernst-effect measurement where a large Nernst signal
indicates the presence of well-defined vortices and therefore a
phase coherence at least in a local scale \cite{Nernst}. The
importance of superconducting phase fluctuations at low doping was
first noted by Emery and Kivelson \cite{EmeryKivelson}, who
conjectured that the whole pseudogap regime might be explained by
a robust pairing amplitude in the presence of strong phase
fluctuations. The problem, however, is that the pairing amplitude
itself is not robust when phase fluctuations are very strong.
Since the electron number is conjugate with the phase, small
fluctuations in electron number at low doping give rise to large
fluctuations in phase. So, while the pseudogap associated with the
singlet formation remains finite even at low doping, the one
associated with the local phase coherence vanishes, as seen
experimentally.

Therefore, it is important to distinguish the above two pseudogap
phenomena: one associated with the singlet formation  and the
other with the local phase coherence. On the other hand, the real
superconducting gap is determined from the formation of sharp
coherence peaks in the density of states. In technical terms, the
real superconducting gap is identified with the fully renormalized
pairing amplitude while the pseudogap associated with the singlet
formation is the bare pairing amplitude. From this view, the
pseudogap associated with local phase coherence is a crossover
energy scale for a partial coherence. In this paper, we would like
to provide a theoretical framework which addresses these issues in
a computationally reliable manner.

To this end, the rest of the paper is organized as follows. We
begin in Section \ref{Hamiltonian} by providing a physical
motivation for the main Hamiltonian of the paper, i.e., the BCS
Hamiltonian with on-site repulsive interaction, $U$, which becomes
identical to the Gutzwiller-projected BCS Hamiltonian in the large
$U$ limit. We call this Hamiltonian the BCS+$U$ Hamiltonian. In
Section \ref{DMFTformalism} the dynamical mean field theory (DMFT)
is formulated for the analysis of the BCS+$U$ Hamiltonian in order
to study $d$-wave superconducting fluctuations under the influence
of $U$. Our DMFT formalism maps the full lattice model to an
effective model for an impurity submerged in $d$-wave
superconducting media. In Section \ref{singlesiteDMFT}, we provide
an argument for the applicability of the single-site DMFT
formalism for the description of superconducting fluctuations near
the insulating phase transition. Note that such situation can be
obtained, for example, at sufficiently low doping close to the
N\'{e}el state, which is actually the main region of interest in
this work. It is further supported by large-scale exact
diagonalization that the assumption of the single-site DMFT
formalism can be in fact valid even away from the phase transition
point as long as the $d$-wave pairing amplitude is appropriately
chosen in the bare level. Detailed derivation of the DMFT
self-consistency equations is presented in Section
\ref{DMFTdetails}.

Main results are reported in Section \ref{results}. First, in
Section \ref{localsimilarity}, we show that, at half filling,
there is a close similarity between the local physics of the
Hubbard model and the strongly-paired BCS+$U$ model, which is
fundamentally connected to the precise equivalence between the
Heisenberg model and the strongly-paired Gutzwiller-projected BCS
model \cite{Park}. We then show in Section \ref{SIT} that the
superconductor-to-insulator transition is mainly caused by the
collapse of the quasiparticle spectral weight (or the $Z$ factor)
which affects both the quasiparticle effective mass and the
superconducting amplitude equally. We emphasize our viewpoint
that, while the real pairing amplitude is fully renormalized to be
zero, the bare pairing amplitude remains non-zero and can be
measured in experiments probing local pairing. In Section
\ref{localphase} we propose a novel method of computing the local
phase stiffness within the DMFT framework. Results from this
method explicitly demonstrate that the phase coherence can survive
locally even after the global coherence is lost. In Section
\ref{doping} we study the doping dependence of the local density
of states (LDOS) which shows that hidden superconductivity
reemerges upon doping. The paper is finally concluded in Sec.
\ref{conclusion} where our theory is placed in perspective.

\section{BCS+$U$ Hamiltonian and the Gutzwiller projection}
\label{Hamiltonian}

Our theory is based on the analysis of the Gutzwiller-projected
BCS Hamiltonian \cite{Park}. In particular, we study the $d$-wave
BCS Hamiltonian with on-site repulsive interaction, $U$, which
plays a role of physically imposing the partial Gutzwiller
projection:
\begin{eqnarray}
H_{\textrm{BCS}+U}&=&-t \sum_{\langle i,j \rangle} (
c^{\dagger}_{i\sigma} c_{j\sigma} + \textrm{H.c.}) -\mu\sum_i
c^{\dagger}_{i\sigma} c_{i\sigma}
\nonumber \\
&+&\sum_{\langle i,j \rangle} \Delta_{ij} (c^{\dagger}_{i\uparrow}
c^{\dagger}_{j\downarrow} -c^{\dagger}_{i\downarrow}
c^{\dagger}_{j\uparrow} + \textrm{H.c.})
\nonumber \\&+&U\sum_{i}
n_{i\uparrow} n_{i\downarrow} \;, \label{BCS+U}
\end{eqnarray}
where the spin index $\sigma = \uparrow$ or $\downarrow$, $t$ is
the hopping amplitude, and $\mu$ is the chemical potential.
$\langle i,j \rangle$ indicates that $i$ and $j$ are nearest
neighbors. Since we are interested in $d$-wave pairing, the bare
pairing amplitude $\Delta_{ij}=\Delta$ for $j=i+\hat{x}$ and
$-\Delta$ for $j=i+\hat{y}$. As mentioned previously, we call this
model the BCS+$U$ model. The large $U$ limit corresponds to the
fully Gutzwiller-projected BCS Hamiltonian.

The main reason for analyzing the Gutzwiller-projected BCS
Hamiltonian is the existence of an exact mapping to the Heisenberg
model at half filling and generally a very close connection to the
$t$-$J$ model at moderate doping, as shown by exact
diagonalization as well as analytic treatments \cite{Park}. We
emphasize that, while the Gutzwiller-projected BCS wave function
[equivalently the resonating valence bond (RVB) state
\cite{Anderson}] is not a good wave function at half filling and
presumably so at low enough doping, the Gutzwiller-projected BCS
Hamiltonian itself serves as a reliable model. Also in a purely
phenomenological level the Gutzwiller-projected BCS Hamiltonian
can be taken as a good theoretical model since superconductivity
in cuprates actually coexists with strong on-site repulsion.

\section{Formulation of the dynamical mean field theory}
\label{DMFTformalism}

For concrete analysis, we develop the dynamical mean field theory
(DMFT) for the BCS+$U$ model. In essence, the dynamical mean field
theory is a quantum generalization of the classical mean field
theory with a key difference that it averages out only spatial
variations while fully taking into account quantum-mechanical,
temporal fluctuations \cite{DMFT_review}. For this reason, the
dynamical mean field theory is regarded as one of the most
powerful theoretical tools in attacking strongly correlated
electron problems, provided that spatial fluctuations are not
strong enough to significantly modify the bare dispersion. We
choose to use the dynamical mean field theory for the analysis of
the BCS+$U$ model since it is known that (i) the Fermi surface as
well as the gap structure are well described by the bare
dispersion of the $d$-wave BCS Hamiltonian, and (ii) the pseudogap
also has a similar $d$-wave structure as the real superconducting
gap \cite{Timusk}.

Actually, there have already been previous attempts to use the
dynamical mean field theory to investigate the existence of
superconductivity in the Hubbard model. The basic idea was to
incorporate the superconducting order parameter via the Nambu
spinor formalism \cite{DMFT_review}. A more recent development was
to combine the Nambu spinor formalism and the cellular dynamical
mean field theory (CDMFT) which enlarges the DMFT impurity from a
single site to a cluster \cite{KyungTremblay,Capone}. An important
advantage of this approach is the flexibility to allow non-local
pairing such as $d$-wave since now electrons can be paired with
those in other sites within cluster. Our approach is completely
different since we first take the Gutzwiller-projected BCS
Hamiltonian to be derived from a more fundamental model such as
the $t$-$J$ model \cite{Park}. Knowing that the full Gutzwiller
projection is obtained in the large $U$ limit, we then analyze the
BCS+$U$ Hamiltonian as a function of $U$ via single-site DMFT
techniques.  The non-local nature of $d$-wave pairing is embedded
in the dispersion of the bare pairing amplitude.

In the following section, we provide an argument that, near the
insulating phase transition (for example, at sufficiently low
doping), the $d$-wave pairing term present in the bare level is
qualitatively sufficient to capture the essence of superconducting
fluctuations. It is further supported by exact diagonalization
that the assumption of the single-site DMFT formalism is in fact
valid even away from the phase transition point as long as the
bare $d$-wave pairing amplitude is appropriately chosen.

\subsection{Applicability of the single-site DMFT formalism}
\label{singlesiteDMFT}


Let us consider the most general form of the Green's function:
\begin{eqnarray}
\hat{G}_{\bf k}^{-1}(i\omega)&\equiv&\left(
\begin{array}{cc}
G_{\bf k}(i\omega) & F_{\bf k}(i\omega) \\
F^*_{\bf k}(i\omega) & -G^*_{\bf k}(i\omega)
\end{array}
\right)^{-1} \nonumber \\
&=&\left(
\begin{array}{cc}
i\omega-\xi_{\bf k} & \Delta_{\bf k} \\
\Delta_{\bf k} & i\omega+\xi_{\bf k}
\end{array}
\right) \nonumber \\
&-&\left(
\begin{array}{cc}
\Sigma_{\bf k}(i\omega) & S_{\bf k}(i\omega) \\
S^*_{\bf k}(i\omega) & -\Sigma^*_{\bf k}(i\omega)
\end{array}
\right)
\end{eqnarray}
where $\xi_{\bf k}=\varepsilon_{\bf k}-\mu$. Simple matrix
inversion then leads to the following expressions:
\begin{eqnarray}
G_{\bf k}(i\omega)=\frac{-[i\omega+\xi_{\bf k}+\Sigma^*_{\bf
k}(i\omega)]}{|i\omega-\xi_{\bf k}-\Sigma_{\bf
k}(i\omega)|^2+|\Delta_{\bf k}-S_{\bf k}(i\omega)|^2}, \nonumber
\\
F_{\bf k}(i\omega)=\frac{\Delta_{\bf k}-S_{\bf
k}(i\omega)}{|i\omega-\xi_{\bf k}-\Sigma_{\bf
k}(i\omega)|^2+|\Delta_{\bf k}-S_{\bf k}(i\omega)|^2}.
\end{eqnarray}

Now, assuming that the pairing symmetry remains precisely $d$-wave
even after $U$, one can generally write that
\begin{eqnarray}
S_{\bf
k}(i\omega)=\tilde{S}_\textrm{odd}(i\omega)
+\tilde{S}_\textrm{even}(i\omega)\frac{\Delta_{\bf
k}}{\Delta}
\end{eqnarray}
where the momentum-independent components,
$\tilde{S}_\textrm{even}(i\omega)$ and
$\tilde{S}_\textrm{odd}(i\omega)$, denote even and odd functions
of $i\omega$, respectively. Since
$\tilde{S}_\textrm{odd}(i\omega)$ is a local quantity, it can be
computed in the single-site DMFT framework. As presented in Sec.
\ref{results}, however, our single-site DMFT study shows that
$\tilde{S}_\textrm{odd}(i\omega)$ is actually zero to within
numerical accuracy, eliminating the possibility of odd-frequency
pairing. On the other hand, the even-frequency pairing component,
$\tilde{S}_\textrm{even}(i\omega)$, can be expanded in low energy
with even powers of $\omega$: $\tilde{S}_\textrm{even}(i\omega)
\simeq \tilde{S}_0 +\tilde{S}_2 \omega^2 +\cdots$.

Based on good agreement between the Fermi surface shape obtained
from bare hopping parameters (for example, determined in the
first-principle calculations) and that of experiment, it can be
argued that the leading term of the normal self-energy correction
is mostly momentum-independent. In this case, the normal
self-energy correction can be computed within the single-site DMFT
framework. Explicit numerical computation of our DMFT study shows
that $\Sigma(i\omega) \simeq (1-1/Z)i\omega$ in the low-energy
limit (Note that the constant term is not explicitly written since
it simply shifts the chemical potential and so can be absorbed
into $\xi_{\bf k}$). The factor $Z$ is called the quasiparticle
spectral weight. After analytic continuation of $i\omega
\rightarrow \omega +i\delta$, the pole of the Green's function is
given in the low-energy limit by the following equation:
\begin{eqnarray}
(\omega/Z)^2-\xi_{\bf k}^2-\Delta_{\bf
k}^2[1-\tilde{S}_\textrm{even}(\omega)/\Delta]^2=0 \;,
\end{eqnarray}
which can be further reduced by expanding
$\tilde{S}_\textrm{even}(\omega)$ as follows:
\begin{align}
\left[\frac{1}{Z^2}-2\frac{\tilde{S}_2}{\Delta}
\left(1-\frac{\tilde{S}_0}{\Delta}\right)\Delta_{\bf k}^2
\right]\omega^2 =\xi_{\bf k}^2+\Delta_{\bf k}^2 \left(
1-\frac{\tilde{S_0}}{\Delta} \right)^2.
\end{align}
Thus, in general, the momentum dependence of the pole can be
somewhat different from that of the simple single-site DMFT
formalism. However, near the insulating phase transition point
where $Z$ vanishes and so one can ignore the second term in the
left hand side of the pole equation (unless $\tilde{S}_2$ also
diverges), the pole structure is basically identical to the simple
singlet-site DMFT result: $\omega = \pm Z \sqrt{\xi^2_{\bf
k}+\Delta^2_{\bf k}(1-\tilde{S}_0/\Delta)^2}$. The only difference
is that the magnitude of the bare pairing amplitude is redefined.
Thus, in this regime, it is qualitatively sufficient to keep track
of only the frequency dependency of the normal self-energy
correction (with the bare pairing amplitude redefined), which is
precisely the situation where the single-site DMFT formulation is
valid.

Actually, we can make a stronger statement based on a large-scale
exact diagonalization study. Poilblanc and Scalapino performed
exact diagonalization of the $t$-$J$ Hamiltonian on a 32-site
square lattice with two doped holes \cite{PS}. In this study,
$G_{\bf k}(\omega+i\delta)$ and $F_{\bf k}(\omega+i\delta)$ were
explicitly computed by using the well-known continued-fraction
method and its extension devised by Ohta and his collaborators
\cite{Ohta}. By assuming the generic form of the Green's
functions, the frequency-dependent gap function, $\Delta_{\bf
k}(\omega)$, was directly computed without any fitting procedures.
The main conclusion is that $\Delta_{\bf k}(\omega)$ has $d$-wave
pairing symmetry and, most importantly, is real and {\it essential
constant} over an energy region larger than the gap itself. This
means in our language that, even if it exists, the anomalous
self-energy correction is essentially a constant (i.e.,
$\tilde{S}_2=0$) over a reasonably wide range of energy, which can
be absorbed into the bare $d$-wave pairing amplitude. It is
important to note that this supports the validity of the
single-site DMFT formalism regardless of the value of the $Z$
factor.

\subsection{Effective impurity-bath Hamiltonian in superconducting media}
 \label{DMFTdetails}

We begin our quantitative analysis by writing the effective
impurity-bath Hamiltonian for the BCS+$U$ model:
\begin{eqnarray}
H_{\textrm{i-b}} &=& \varepsilon_c c^{\dagger}_{\sigma} c_{\sigma}
+U n_{c\uparrow} n_{c\downarrow}
\nonumber \\
&+& \sum_l (
\begin{array}{cc}
a^{\dagger}_{l\uparrow} & a_{l\downarrow}
\end{array}
)
\left(
\begin{array}{cc}
\tilde{\varepsilon}_l & \tilde{\Delta}_l \\
\tilde{\Delta}_l   & -\tilde{\varepsilon}_l
\end{array}
\right)
\left(
\begin{array}{c}
a_{l\uparrow} \\
a^{\dagger}_{l\downarrow}
\end{array}
\right)
\nonumber \\
&+&\sum_l V_l ( a^{\dagger}_{l\sigma} c_{\sigma} +
c^{\dagger}_{\sigma} a_{l\sigma} )
\nonumber \\
&+&\sum_l W_l ( a^{\dagger}_{l\uparrow} c^{\dagger}_{\downarrow} -
a^{\dagger}_{l\downarrow} c^{\dagger}_{l\uparrow} +\textrm{H.c.}),
\label{Hib}
\end{eqnarray}
where $c_\sigma$ and $a_{l\sigma}$ are, respectively, the
operators for impurity and bath orbitals. The core energy,
$\varepsilon_c$, is minus the chemical potential. $V_l$ is the
usual hybridization parameter for hopping between impurity and
bath, while $\tilde{\varepsilon}_l$ is the energy of the $l$-th
bath orbital. Crucial additions in our theory are
$\tilde{\Delta}_l$, the pairing amplitude of the $l$-th bath
orbital, and $W_l$, the anomalous hybridization parameter for
pairing between impurity and bath.

To insure that $H_\textrm{i-b}$ is a faithful effective
Hamiltonian for the BCS+$U$ model, unknown effective parameters,
$\tilde{\varepsilon}_l$, $\tilde{\Delta}_l$, $V_l$, and $W_l$,
should be determined by the DMFT self-consistency equation which
imposes the condition that the impurity Green's function is
entirely equivalent to the local Green's function of the lattice:
\begin{eqnarray}
\hat{G}(i\omega) = \sum_{\bf k} \hat{G}_{\bf k}(i\omega) \;,
\label{SCE}
\end{eqnarray}
where
\begin{eqnarray}
\hat{G}_{\bf k}^{-1}(i\omega)=\left(
\begin{array}{cc}
i\omega-\varepsilon_c-\varepsilon_{\bf k} & \Delta_{\bf k} \\
\Delta_{\bf k} & i\omega+\varepsilon_c+\varepsilon_{\bf k}
\end{array}
\right) -\hat{\Sigma}(i\omega) \;,
\label{Gk}
\end{eqnarray}
in which $\varepsilon_{\bf k}$ and $\Delta_{\bf k}$ are given by
the bare dispersion; $\varepsilon_{\bf k} =
-2t(\cos{k_x}+\cos{k_y})$ and $\Delta_{\bf k} =
2\Delta(\cos{k_x}-\cos{k_y})$.

The self-energy correction, $\hat{\Sigma}(i\omega)$, is the
difference between the inverse of the impurity Green's function
for $U=0$ (called the Weiss field), $ \hat{\cal
G}_0^{-1}(i\omega)$, and the inverse of the full impurity Green's
function for finite $U$, $\hat{G}^{-1}(i\omega)$:
\begin{eqnarray}
\hat{\Sigma}(i\omega)= \hat{\cal G}_0^{-1}(i\omega) -
\hat{G}^{-1}(i\omega) \;,
\end{eqnarray}
where
\begin{eqnarray}
\hat{G}(i\omega) = \left(
\begin{array}{cc}
G(i\omega) & F(i\omega) \\
F(i\omega) & -G^*(i\omega)
\end{array}
\right) \;,
\end{eqnarray}
in which $G(i\omega)$ and $F(i\omega)$ are, respectively, the
normal and anomalous impurity Green's function.


The Weiss field, $\hat{\cal G}_0(i\omega)$, can be in turn
expressed in terms of the effective parameters,
$\tilde{\varepsilon}_l$, $\tilde{\Delta}_l$, $V_l$, and $W_l$:
\begin{align}
\hat{\cal G}^{-1}_0(i\omega) =
 \left(
\begin{array}{cc}
i\omega-\varepsilon_c & 0 \\
0 & i\omega+\varepsilon_c
\end{array}
\right) -\left(
\begin{array}{cc}
\Gamma(i\omega) & \Lambda(i\omega) \\
\Lambda(i\omega) & -\Gamma^*(i\omega)
\end{array}
\right) ,
\label{Gw}
\end{align}
where $\Gamma(i\omega)$ and $\Lambda(i\omega)$ are the normal and
anomalous corrections in the impurity Green's function after
integrating out all bath orbitals. Explicitly,
\begin{align}
\Gamma(i\omega)=&\sum_l \Big( V^2_l G_l-W^2_l G^*_l +2V_l W_l F_l
\Big),
\nonumber \\
\Lambda(i\omega)=&\sum_l \Big[(V^2_l-W^2_l) F_l-V_l W_l
(G_l+G^*_l) \Big],
 \label{GammaLambda}
\end{align}
where $G_l$ and $F_l$ are, respectively, the normal and anomalous
Green's functions of the $l$-th bath orbital:
\begin{eqnarray}
G_l(i\omega)&=& \frac{i\omega+\tilde{\varepsilon}_l}
{(i\omega)^2-\tilde{\varepsilon}^2_l-\tilde{\Delta}^2_l} \;,
\nonumber \\
F_l(i\omega)&=&
\frac{-\tilde{\Delta}_l}{(i\omega)^2-\tilde{\varepsilon}^2_l-\tilde{\Delta}^2_l}
\;.
\end{eqnarray}

Solving the DMFT self-consistency equation in Eq.(\ref{SCE})
requires the full impurity Green's function for a given set of the
effective parameters, which is computed in our study via exact
diagonalization with a finite number of orbitals which is ten
throughout this paper. Computing the full impurity Green's
function is technically difficult and numerically expensive in
general, but is particularly time-consuming for our study because
$H_{\textrm{i-b}}$ does not conserve the particle number so that
different number-sectors mix and $H_{\textrm{i-b}}$ needs to be
diagonalized in the Hilbert space containing all possible number
configurations.
Once the impurity Green's function is known, the effective
parameters in $H_\textrm{i-b}$ can be determined iteratively in a
similar manner to Caffarel and Krauth \cite{CaffarelKrauth}. The
full impurity Green's function obtained in the converged iteration
is taken to be the final solution.


Before moving to results reported in the next section, the
following technical aspect regarding the convergence of effective
parameters is noteworthy. While initial values for the effective
parameters can be in principle chosen arbitrarily, it is much
convenient to use symmetry constraints in order to reduce any
unnecessary degrees of freedom. In our case, the $d$-wave pairing
symmetry imposes a constraint on the effective parameters so that
two identical parameter sets for $\tilde{\varepsilon}_l$, $V_l$,
and $W_l$ always come in pair with the opposite sign of
$\tilde{\Delta}_l$, i.e., $(\tilde{\varepsilon}_l,
\tilde{\Delta}_l, V_l, W_l)$ and $(\tilde{\varepsilon}_l,
-\tilde{\Delta}_l, V_l, W_l)$. At half filling, there is an
additional constraint due to the particle-hole symmetry so that,
for a given set
$(\tilde{\varepsilon}_l,\tilde{\Delta}_l,V_l,W_l)$, there exists a
corresponding set of
$(-\tilde{\varepsilon}_l,\tilde{\Delta}_l,V_l,W_l)$.


\section{Results}
\label{results}

\subsection{Local similarity between the Hubbard model
and the strongly-paired BCS+$U$ model}
\label{localsimilarity}

\begin{figure}
\includegraphics[width=2.4in]{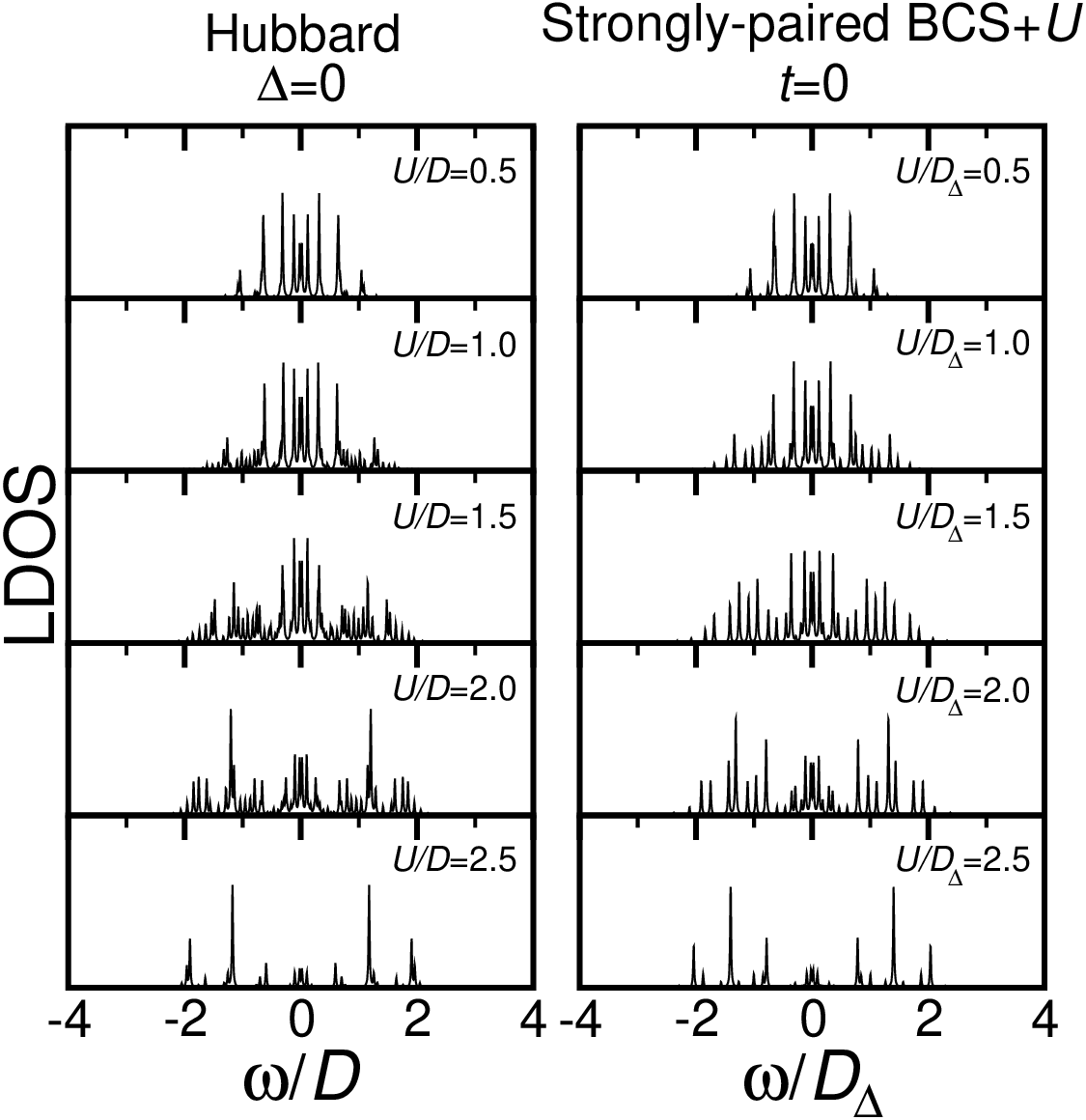}
\caption{Local density of states (LDOS) for the Hubbard model
($\Delta=0$) and the strongly-paired BCS+$U$ model ($t=0$) at half
filling as a function of on-site repulsive interaction, $U$. In
the plot, $D=4t$ and $D_\Delta=4\Delta$.
\label{fig1}}
\end{figure}

For results, we first compute the local density of states (LDOS)
for the strongly-paired BCS+$U$ model with $\varepsilon_{\bf
k}=0$, which is a meaningful model at half filling since, in the
large-$U$ limit, this model is precisely identical to the
Heisenberg model \cite{Park}. Figure~\ref{fig1} presents the
comparison between the LDOS for this model and that for the
Hubbard model (i.e., $\Delta_{\bf k}=0$), which shows that the two
models have the almost identical {\it local} physics despite the
difference that, at small $U$, the former has superconductivity
and the latter does not. In fact, this local similarity is rather
important and assuring since it suggests that the DMFT framework
can capture not only the fundamental equivalence between the two
models in the large $U$ limit, but also their close relationship
at general $U$. It is emphasized that, as a function of $U$, the
pairing term alone generates the self-energy correction
practically identical to that of the Hubbard model.

Two facts are important to understand the nature of the insulating
transition at large $U$. First, it is shown in our DMFT
calculation that the anomalous self-energy correction in fact
vanishes, eliminating the possibility of odd-frequency pairing.
While at half filling this fact can be deduced rather
straightforwardly by using (i) the particle-hole symmetry and (ii)
the $d$-wave pairing symmetry, the situation is less clear at
finite doping. It is only through explicit numerical computation
that the anomalous self-energy correction is shown to become zero
due to intricate adjustment of relevant parameters in the
effective impurity-bath Hamiltonian in Eq.~(\ref{Hib}).

Second, $\Sigma(i\omega)$ becomes a linear function of $i\omega$
in the low-energy limit, at least, within the metallic regime:
$\Sigma \simeq U/2+(1-1/Z) i\omega$. Note that the constant term
of the normal self-energy exactly cancels the chemical potential
at half filling. In this case the denominator of the normal
Green's function simply becomes $\omega^2/Z^2-\Delta^2_{\bf k}$
after the analytic continuation of $i\omega \rightarrow \omega
+i\delta$ since the denominator of $G(i\omega)$ is $|i\omega
-\varepsilon_c -\Sigma(i\omega)|^2 +\Delta^2_{\bf k}$ when the
anomalous self-energy correction is zero. In the above, the core
energy, $\varepsilon_c$ $(= -\mu)$, is $-U/2$ at half filling. The
pole structure at $\omega=\pm Z |\Delta_{\bf k}|$ indicates that
the pairing dispersion is effectively renormalized to be $Z
\Delta_{\bf k}$. As implied in Fig.~\ref{fig1}, $Z$ decreases as a
function of $U$ and finally vanishes at a critical value, $U_c$,
inducing the collapse of superconductivity. Therefore, in our DMFT
framework, $Z$ contains the major effect of superconducting
fluctuations.

It is interesting to note that the strongly-paired BCS+$U$ model
shows no gap structure in the local density of states. Instead,
this model exhibits the characteristic Kondo resonance peak
similar to that of the Hubbard model. We emphasize that in the
above $\Sigma(i\omega)$ is generated solely from the $d$-wave
pairing term in the presence of $U$. In the general BCS+$U$ model
where both $t$ and $\Delta$ are not zero, the self-energy
corrections from the both terms mix, which in turn shifts $U_c$
from that of the pure Hubbard model (or equivalently that of the
strongly-paired BCS+$U$ model). Also, the existence of finite
$\Delta/t$ shows up as sharp superconducting coherence peaks in
the local density of states. We discuss this problem in the next
section.

\subsection{Superconductor-to-insulator transition in the BCS+$U$ model}
\label{SIT}

\begin{figure}
\includegraphics[width=2.7in]{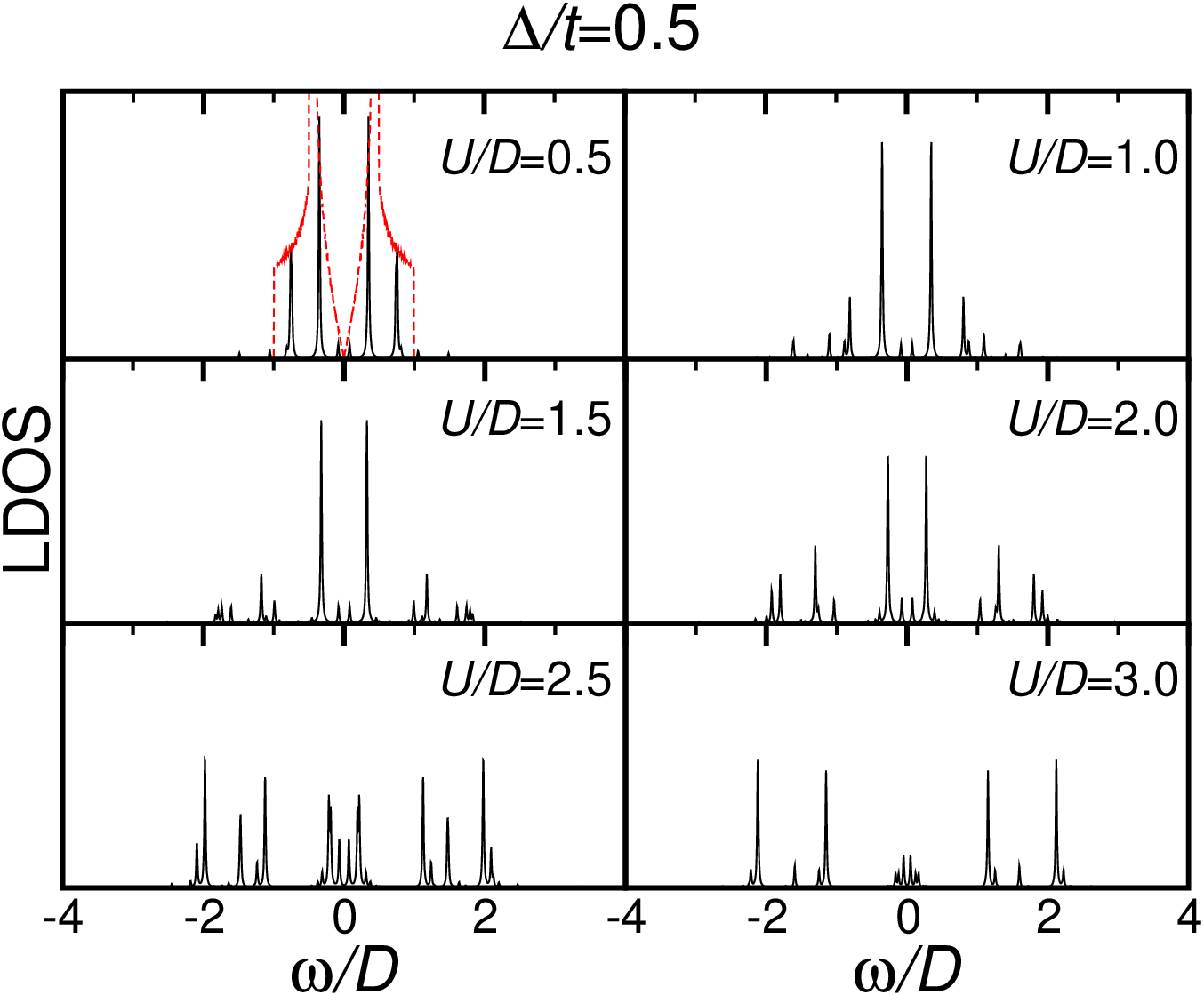}
\caption{Local density of states (LDOS) for the BCS+$U$ model with
$\Delta/t=0.5$ at half filling as a function of $U$. $\Delta$
denotes the bare pairing amplitude and $D (=4t)$ indicates half
the bare band width. The dotted curve in the top, left panel
indicates the LDOS analytically obtained for the non-interacting
system with $U/D=0$. \label{fig2}}
\end{figure}
Figure \ref{fig2} plots the LDOS for the BCS+$U$ Hamiltonian with
$\Delta/t=0.5$ at half filling as a function of $U$. Due to
pairing, the density of states is reduced in the vicinity of the
Fermi level (i.e., near $\omega/D=0$ in the plot) and the
coherence peak develops at the position of basically the
superconducting energy gap. What is most important in
Fig.~\ref{fig2} (and also in Fig.~\ref{fig4}) is that the
superconducting gap closes as $U$ increases, and eventually
disappears for $U/D \gtrsim 3.5$. (Note that, despite its discrete
nature, the highest LDOS peak near the Fermi level can be taken to
be a good indicator for the real coherence peak in the continuum
limit, as demonstrated in the top, left panel of Fig.~\ref{fig2}.)
It is important to note that the bare pairing amplitude is set to
be constant. Since the bare pairing amplitude is physically
equivalent to the energy gap for singlet formation
\cite{Park,PRT,ALRTZ}, the large renormalization of the bare
pairing amplitude provides an explanation for the pseudogap
phenomenon. That is to say, at half filling, the real
superconducting gap is strongly renormalized from the bare pairing
amplitude and finally becomes zero in the large $U$ limit. So,
superconductivity is hidden at half filling while singlet pairs
are inherently present due to the non-zero bare pairing amplitude.


\begin{figure}
\includegraphics[width=2.1in]{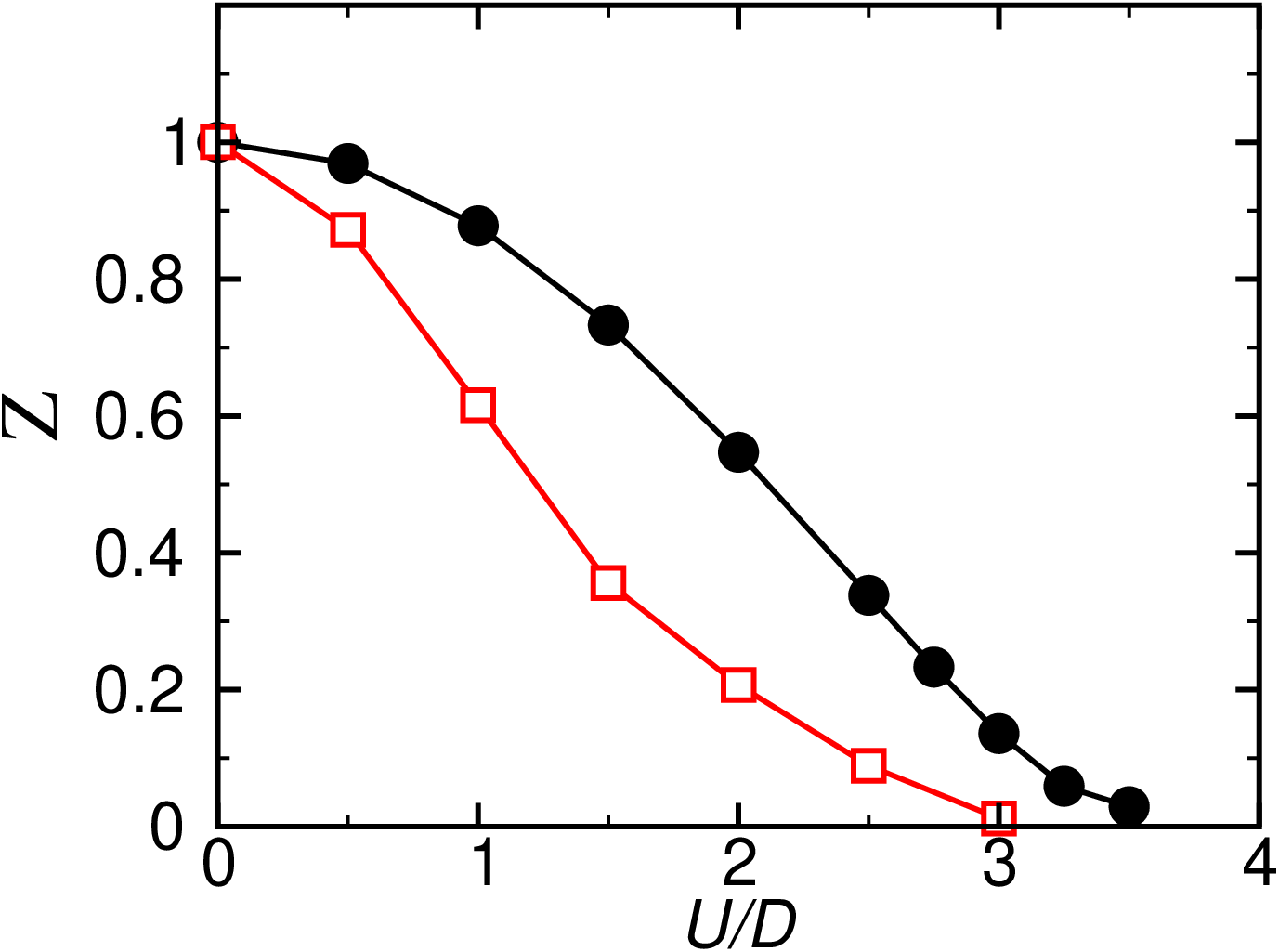}
\caption{Quasiparticle spectral weight, Z, as a function of $U$.
Solid circles indicate $Z$ of the BCS+$U$ model with
$\Delta/t=0.5$ while empty squares denote that of the Hubbard
model. The error bar due to the fitting procedure necessary for
determining $Z$ from the low-energy behavior of $\Sigma(i\omega)$
is estimated to be smaller than the size of the symbol. Note that,
since the above calculation is performed at zero temperature, we
ignore the second DMFT solution which is adiabatically connected
to the insulating phase generating smaller $Z$ \cite{DMFT_review}.
At zero temperature this second solution is higher in energy than
the metallic (superconducting) solution. \label{fig3}}
\end{figure}
As mentioned in the previous section, the
superconductor-to-insulator transition is driven by the collapse
of the quasiparticle spectral wight, $Z$. It is shown also in the
previous section that $Z$ of the strongly-paired BCS+$U$ model
(i.e., when $\Delta/t \rightarrow \infty$) has the $U$ dependence
almost identical to that of the pure Hubbard model. When
$\Delta/t$ is finite, however, the self-energy corrections from
the hopping and pairing term mix and thus $Z$ is modified.
Figure~\ref{fig3} plots $Z$ of the BCS+$U$ model with
$\Delta/t=0.5$ in comparison to that of the Hubbard model. As seen
in the plot, $Z$ is modified so that it is enhanced for a given
$U$, leading to an increased $U_c$. It is interesting to observe
that, when coexisting with the hopping term, the pairing
correlation reduces the effect of $U$ and delays the insulating
phase transition.

It has been shown so far in this section that superconductivity is
suppressed for sufficiently large $U$ at half filling. A natural
question that follows is if superconductivity can actually
reemerge when electrons become mobile upon doping. Before
addressing this issue in Sec.~\ref{doping}, however, we first
investigate the second kind of the pseudogap phenomenon which is
associated with the local phase coherence.

\subsection{Local phase stiffness and the pseudogap} \label{localphase}

We first need to quantify the local phase coherence. Fortunately,
our DMFT formalism renders a very natural way to measure the local
phase coherence by implementing a twist in local phase. The
superconducting phase of a locally coherent region (which includes
a given site and its correlated surroundings) can be twisted by
performing $W_l \rightarrow W_l e^{i\theta}$ for every $l$, while
the superconducting phase of the bath, $\tilde{\Delta}_l$, remains
fixed. In other words, the anomalous hybridization between a given
site and every bath orbital connected to it is twisted in phase.

\begin{figure}
\includegraphics[width=2.2in]{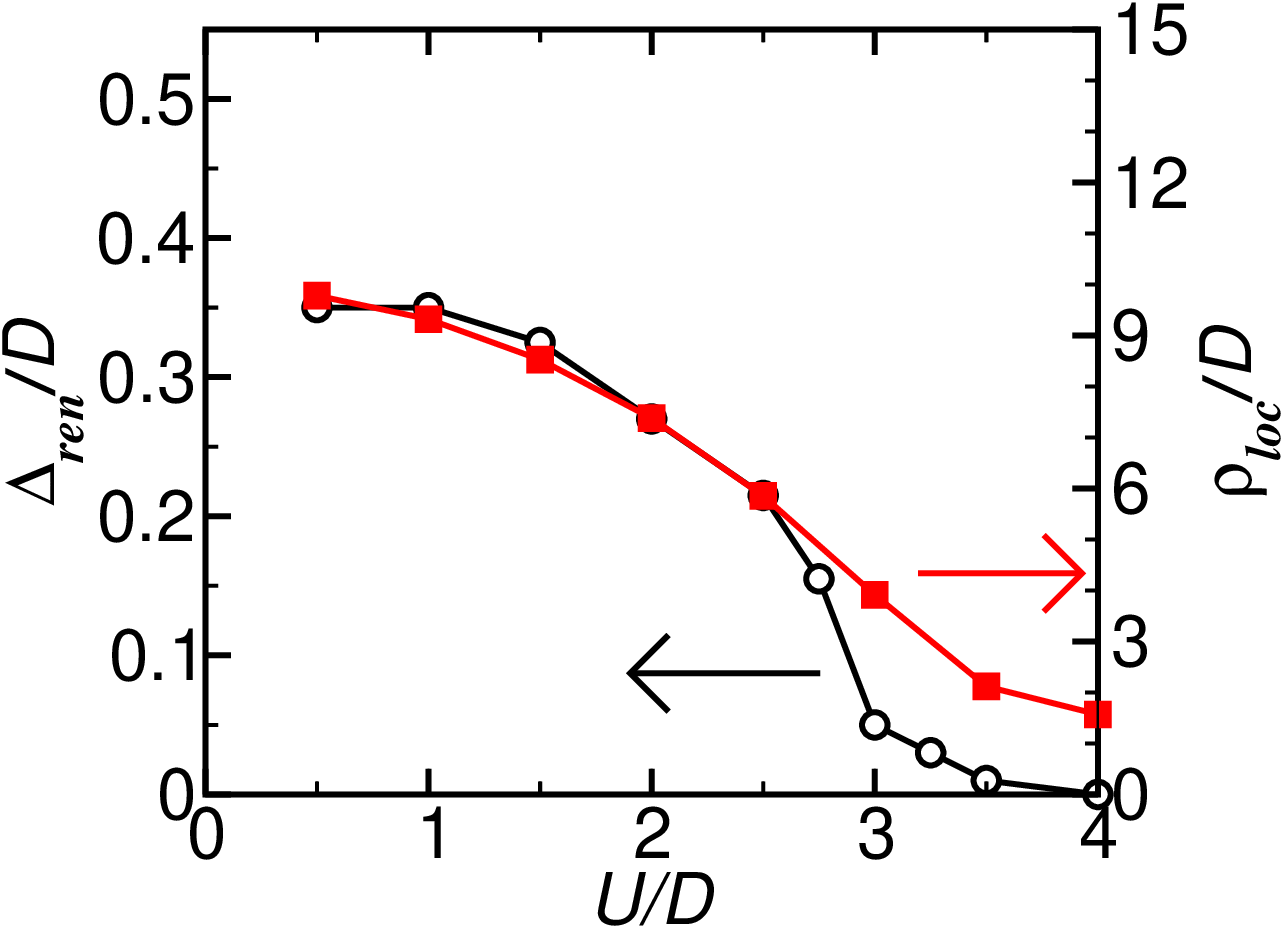}
\caption{$U$ dependence of the renormalized superconducting gap,
$\Delta_\textrm{ren}$ (empty circles), and the local phase
stiffness, $\rho_\textrm{loc}$ (solid squares), at half filling.
$\Delta_\textrm{ren}$ is determined to be the position of the
coherence peak in the local density of states. $\rho_\textrm{loc}$
is computed from the energy cost of local phase twist. Note that
the bare pairing amplitude is constant: $\Delta/t=0.5$.
\label{fig4}}
\end{figure}
To see what this phase twist entails, let us consider its effects
on the Weiss field in Eq.~(\ref{Gw}), which describes the dynamics
of the impurity electron after integrating out the bath. In
particular, $\Gamma(i\omega)$ and $\Lambda(i\omega)$ in
Eq.~(\ref{GammaLambda}) are transformed as follows:
\begin{align}
\Gamma(i\omega)=&\sum_l \Big( V^2_l G_l-W^2_l G^*_l +2 V_l W_l F_l
\cos{\theta} \Big),
\nonumber \\
\Lambda(i\omega)=&\sum_l \Big[(V^2_l-W^2_l e^{2i\theta}) F_l
-V_l W_l e^{i\theta} (G_l+G^*_l) \Big].
\end{align}
Apparently, the effective impurity action is not generally
invariant under the local phase twist. In the special case of the
$d$-wave superconducting bath, however, it turns out that the
impurity action is actually invariant owing to the fact that
$\sum_l V_l W_l F_l=0$ and $\Lambda(i\omega)=0$ for all $\omega$.
Note that $\sum_l V_l W_l F_l=0$ due to the $d$-wave pairing
symmetry and $\Lambda(i\omega)=0$ due to the absence of
odd-frequency pairing. Consequently, eigenenergies are also
invariant. However, since the ground state itself is rotated,
there is an energy cost of the rotated ground state against the
original Hamiltonian once the phase rotation symmetry is
spontaneously broken. This energy cost can be computed as follows:
\begin{eqnarray}
\Delta E (\theta) = \langle \Psi_\theta | H_\textrm{i-b} |
\Psi_\theta \rangle -E_0
\end{eqnarray}
where $|\Psi_\theta \rangle$ is the ground state wave function at
the phase twist of $\theta$, $H_\textrm{i-b}$ is the effective
impurity-bath Hamiltonian at $\theta=0$, and $E_0$ is the ground
state energy at $\theta=0$.

The local phase stiffness, $\rho_\textrm{loc}$, is defined as the
curvature of the energy cost with respect to the local phase
twist: $\Delta E \simeq \rho_\textrm{loc} \theta^2 /2$ for
sufficiently small $\theta$. Figure \ref{fig4} shows the $U$
dependence of $\rho_\textrm{loc}$ in comparison with that of the
renormalized superconducting gap, $\Delta_\textrm{ren}$, which is
defined as the position of the coherence peak in Fig. \ref{fig2}.
It is important to note that, for $U/D \gtrsim 3.5$, the
superconducting coherence peak completely disappears while the
local phase stiffness remains finite. This disparity is the
essence of the pseudogap phenomenon associated with the local
phase coherence.

\subsection{Reemergence of superconductivity at finite doping}
\label{doping}

\begin{figure}
\includegraphics[width=1.8in]{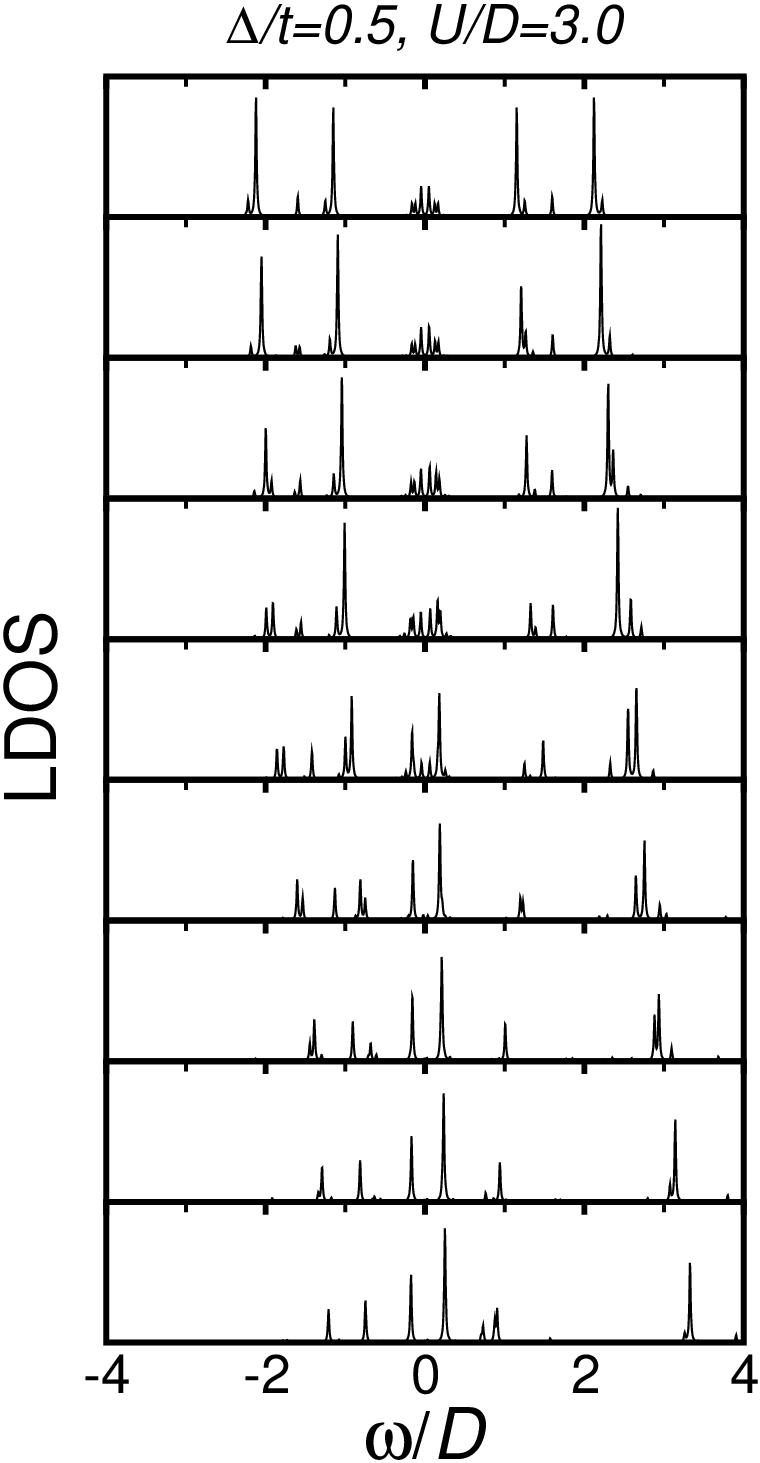}
\caption{Local density of states (LDOS) for the BCS+$U$ model with
$\Delta/t=0.5$ and $U/D=3$ as a function of doping. The value of
$U/D=3$ is chosen so that the ground state at half filling is just
about to enter the insulating phase, which corresponds to the
almost fully Gutzwiller-projected regime. The top panel plots the
LDOS at half filling which is obtained when $\mu= \mu_0 \equiv
U/2$. From the top to bottom panels, the chemical potentials (the
corresponding hole concentration, $x$) are $\mu_0$ ($0\%$),
$0.9\mu_0$ ($1.65\%$), $0.8\mu_0$ ($3.70\%$), $0.7\mu_0$
($6.42\%$), $0.6\mu_0$ ($9.90\%$), $0.5\mu_0$ ($14.10\%$),
$0.4\mu_0$ ($19.30\%$), $0.3\mu_0$ ($25.13\%$), and $0.2\mu_0$
($31.53\%$), respectively. \label{fig5}}
\end{figure}

We now turn to the doped BCS+$U$ model. As demonstrated in
Fig.~\ref{fig4}, the superconducting gap is suppressed at half
filling for sufficiently large $U/D$. The question is whether
hidden superconductivity can reemerge when phase fluctuations are
lessened by allowing charge fluctuations via doping.
Figure~\ref{fig5} shows that the increase in hole concentration
indeed gradually opens up the superconducting gap at the Fermi
level of $\omega/D=0$. It is interesting to observe the specific
route in which the superconducting gap is opened up. As the hole
concentration increases, the lower Hubbard band moves toward the
Fermi level and provides a necessary density of states for
suppressed superconductivity to become reactivated.

Note that $\Delta/t=0.5$ is an appropriate value for the $t$-$J$
model in the regime of exchange coupling $J/t \simeq 0.5-1.0$ and
hole concentration $x \simeq 10-15 \%$ (optimal doping). As
mentioned in Sec.~\ref{Hamiltonian}, such connection between the
$t$-$J$ model and the Gutzwiller-projected BCS model is
established via wave function overlap using exact diagonalization
\cite{Park}. Therefore, in Fig.~\ref{fig5} the fifth or sixth
panel from the top corresponds to the actual local density of
states of the $t$-$J$ model in the regime specified in the above.
It will be interesting for future work to conduct a systematic
survey of the renormalized superconducting gap as a function of
doping, obtained from the bare pairing amplitude properly
determined from exact diagonalization.

\section{Conclusion}
\label{conclusion}

In this paper, the dynamical mean field theory is used for the
analysis of the BCS+$U$ Hamiltonian with $U$ imposing the
(partial) Gutzwiller projection. Strong phase fluctuations at half
filling are manifested through the collapse of the quasiparticle
spectral weight, $Z$, which affects not only the quasiparticle
effective mass, but also the pairing amplitude. As a consequence,
the distance between the superconducting coherence peaks in the
density of states (which measures the renormalized superconducting
gap) decreases at the same time as the overall weight is reduced.
It is thus fundamentally due to strong phase fluctuations that the
renormalized superconducting gap decreases as a function of $U$
and finally vanishes for a given finite bare pairing amplitude.

Moreover, it is shown that the local phase coherence can survive
even after the global phase coherence identified with the sharp
superconducting coherence peak is lost. Combined with this result
for the local phase coherence, the collapse of the renormalized
superconducting gap provides an explanation for the pseudogap
phenomenon. That is, there are two different pseudogap energy
scales, one of which is associated with the bare pairing amplitude
that can remain finite even at half filling. The other is
associated with the local phase stiffness that vanishes at
sufficiently small doping in the large $U$ limit. In this view,
the real superconducting gap is given by the fully renormalized
pairing gap which is the smallest energy scale at small doping for
finite $U$ and vanishes faster than the other two energy scales as
$U$ increases.
It is further shown that hidden superconductivity at half filling
can indeed reemerge upon doping, as indicated by the reappearance
of the superconducting coherence peaks in the density of states.

We finally conclude by putting our theory in perspective.
Obviously, it is well beyond the scope of this paper to provide a
complete review of previous theories on the pseudogap phenomena.
Thus, we select only a few theories that are closely related to
our theory. In an overall scheme our theory can be categorized as
the theory with preformed pairs since electron pairs are
inherently present due to the bare pairing amplitude. There are,
however, many variations in the theory of preformed-pair scenario.
Pairs can be preformed by (i) the spin-charge separation
\cite{KotliarLiu,LeeNagaosa,LeeWen} involving deconfined phases
\cite{LeeNagaosaWen}, (ii) the formation of microscopic stripes
\cite{Stripe}, (iii) the proximity to the long-range
antiferromagnetic order \cite{MMP}(or the nearly antiferromagnetic
Fermi liquid theory), and so on. In our theory, preformed pairs
exist because the ground state itself has the bare pairing
amplitude due to the fundamental connection between the $t$-$J$
model and the projected BCS Hamiltonian. The dichotomy between the
bare and real superconducting gap originates from the strong
renormalization effect of large on-site repulsive interaction,
$U$.

It is interesting to note that the above-mentioned theories
including ours can be categorized further into two classes, as
proposed by Lee, Nagaosa, and Wen. In their review paper
\cite{LeeNagaosaWen}, Lee {\it et al.} coined the words, ``thermal
explanation" and ``quantum explanation" of the pseudogap. The
quantum explanation of the pseudogap proposes a fundamentally new
quantum state which, for example, is the deconfined spin-liquid
phase in the above spin-charge separation scenario. Despite the
fact that this deconfined spin-liquid state may be unstable in the
square-lattice $t$-$J$ model, it is assumed that the pseudogap is
a high-frequency property of the spin-liquid phase, seen at high
temperature.

On the other hand, the thermal explanation refers to any theories
viewing the pseudogap as a finite-temperature manifestation of the
spin gap necessary for the formation of singlet pairs, which is
induced fundamentally by the symmetry-breaking ground state at
zero temperature. Broken symmetries are the lattice translation,
spin rotation, and global gauge invariance in the case of the
stripe theory, the nearly antiferromagnetic Fermi liquid theory,
and the fluctuating superconductivity theory
\cite{EmeryKivelson}(which conceptually includes our theory of the
Gutzwiller-projected BCS model), respectively. The main difference
between our theory and the others is that the real superconducting
gap originates from the same spin gap while it is strongly
renormalized due to large $U$. Consequently, in our theory, only
the real superconducting gap is shown in the zero-temperature
density of states while a signature for the pseudogap can be seen
in the finite-temperature counterpart.

\begin{center}
\textbf{Acknowledgments}
\end{center}

The author is grateful to Y. Bang, H.-Y. Choi, G. S. Jeon, and
K.-S. Kim for their insightful comments.


\end{document}